\documentclass[conference]{IEEEtran}

\usepackage{cite}
\usepackage{amsmath,amssymb,amsfonts}
\usepackage{algorithmic}
\usepackage{graphicx}
\usepackage{textcomp}
\usepackage{physics}
\usepackage{hyperref}
\usepackage{xcolor}
\usepackage{float}
\def\BibTeX{{\rm B\kern-.05em{\sc i\kern-.025em b}\kern-.08em
    T\kern-.1667em\lower.7ex\hbox{E}\kern-.125emX}}
\begin{document}

\title{Probability Currents in Out-of-Equilibrium Microwave Circuits}

\author{\IEEEauthorblockN{Alexandre Dumont}
\IEEEauthorblockA{
\textit{Institut Quantique,}\\
\textit{Département de physique,} \\
\textit{Université de Sherbrooke}\\
Sherbrooke, Canada \\
Alexandre.Dumont3@usherbrooke.ca}
\and
\IEEEauthorblockN{Pierre Février}
\IEEEauthorblockA{
\textit{Institut Quantique,}\\
\textit{Département de physique,} \\
\textit{Université de Sherbrooke}\\
Sherbrooke, Canada \\
}
\and
\IEEEauthorblockN{Christian Lupien}
\IEEEauthorblockA{
\textit{Institut Quantique,}\\
\textit{Département de physique,} \\
\textit{Université de Sherbrooke}\\
Sherbrooke, Canada \\
}
\and
\IEEEauthorblockN{Bertrand Reulet}
\IEEEauthorblockA{
\textit{Institut Quantique,}\\
\textit{Département de physique,} \\
\textit{Université de Sherbrooke}\\
Sherbrooke, Canada \\
Bertrand.Reulet@usherbrooke.ca}
}

\maketitle

\begin{abstract}
In this work we reconstruct the probability current in phase space of out-of-equilibrium
microwave circuits. This is achieved by a statistical analysis of short-time correlations in time domain measurements.
It allows us to check locally in phase space the violation of detailed balance or the presence of fluctuation loops.
We present the data analysis methods and experimental results for several microwave circuits driven by two noise sources in the 4-8GHz frequency range.
\end{abstract}


\section{Introduction}
Measuring the probability current in phase 
space~\cite{lecomte_energy_2005,gonzalez_experimental_2019,szabo_probability_2010} is an essential tool in 
the analysis of the thermodynamics of noise driven systems. 
It serves as a basis to test detailed balance violation, existence of fluctuation 
loops~\cite{gonzalez_experimental_2019}, demonstrate brownian gyration\cite{chiang_electrical_2017} 
or calculate entropy production for example~\cite{chiang_entropy_2017}.
Measuring the probability current can be done by calculating time-derivative of measured 
voltages~\cite{gonzalez_experimental_2019}. 
This however requires huge oversampling: for the variation in voltage to be small between two measurements, 
the sampling rate of the detection must be much faster than the bandwidth of the detected signal. 
This is easily achieved in the audio frequency range but is totally impractical with microwaves in the GHz range. 
Another approach consist in constructing probability distributions of successive 
measurements~\cite{chiang_entropy_2017},i.e. build histograms of pairs of successive voltages from which 
transition probabilities can be deduced. 
This is the path we have followed, with time traces of voltage with an analog bandwidth of 8GHz recorded at 
a speed of 32GS/s.

The rest of this paper goes over the theoretical aspects of reconstructing the probability current, 
the experimental setup, the data analysis and the results.

\section{Theory}
\subsection{Probability current}
We consider a circuit in which we measure the time-dependent voltage fluctuations at two different locations $V_{1,2}(t)$, 
yielding a two dimensional phase space where points are represented with coordinates
\begin{align}
    \vec{V} = \begin{pmatrix}V_1\\V_2\end{pmatrix}.
\end{align}
We consider a system in a stationary state and note  $P(\vec{V})$ the probability to measure a 
voltage $\vec{V}$ at any instant $t$. 
Voltages are not measured continuously in time, but regularly with a time step $\tau$. 
We note $P_\tau(\vec V,\vec V')$ the probability that two successive measurements (separated by a time $\tau$) 
give $\vec V$ and $\vec V'$ respectively. 
The transition rate from $\vec V$ to $\vec V'$ is $\Gamma(\vec V,\vec V')=(1/\tau)P_\tau(\vec V,\vec V')/P(\vec V)$. 
The phase space current at point $\vec V$ is:
\begin{align}
    \vec J(\vec{V}) &= \sum_{\vec{V}'} [P_\tau(\vec{V},\vec{V}')-P_\tau(\vec{V}',\vec{V})]\frac{\vec{V}'-\vec{V}}\tau
\end{align}
Introducing the (scalar) probability current between $\vec V$ and $\vec V'$ 
\begin{align}
    j(\vec{V},\vec{V}') &= P(\vec{V})\Gamma(\vec{V},\vec{V}')-P(\vec{V}')\Gamma(\vec{V}',\vec{V}),
    \label{eq:j}
\end{align}
One has: 
\begin{align}
    \vec{J}(\vec{V}) &= \sum_{\vec{V}'}j(\vec{V},\vec{V}')(\vec{V}'-\vec{V}).
    \label{eq:vec_J}
\end{align}

The detailed balance condition is $j(\vec{V},\vec{V}')=0$ for all $\vec{V}$, $\vec{V}'$. In the following we 
show measurements of both $j$ and $\vec J$ in various circuits containing two noise sources with identical or 
different noise temperatures, which correspond respectively to circuits at equilibrium or driven out of 
equilibrium by a temperature difference.
 
\section{Experimental setup}
The experimental setup contains three elements: i) two noise sources that generate voltage fluctuations of 
controllable variance, ii) a microwave circuit that is fed by the noise sources, and iii) a detection chain. 
The complete circuit is shown in Fig.~\ref{fig:setup}.

\subsection{Noise sources}
The experiment has been performed at room temperature. One of the noise source is simply a 50 Ohm resistor 
(i.e., matched to the microwave circuitry) that generates thermal noise with a noise temperature of 
293K~\cite{johnson_thermal_1928,nyquist_thermal_1928}. 
The second is a commercial electronic calibrated noise source with a noise temperature of 900K across 
the 1 to 15 GHz bandwidth. 
Its noise temperature cannot be modified, but by attenuating the noise generated by this source one can achieve 
a lower noise temperature down to 300K. 
As a matter of fact, what matters for the experiment is the variance of voltage fluctuations, which is expressed 
in terms of equivalent noise temperature, and not the thermodynamic temperature itself.

\subsection{Detection chain}
The voltages $V_{1,2}$ are obtained by digitizing at high speed the voltage at two different locations in the 
circuits, after amplification and filtering. 
Note that, as usual with microwaves, the detection is not of high impedance but has an input impedance of $50\Omega$. 
This has two consequences: first, the measured voltage is that of the wave exiting the circuit, not the local voltage 
in the circuit; second, the detection circuit emits noise towards the circuit, which could affect the measurement. 
As we discuss below, this effect does not take place in our setup: the voltage fluctuations we measure are entirely 
due to the noise sources. 
Note however that the amplifiers add noise to the one measured, but the noises of the two amplifiers are uncorrelated.
The digitization of the two signals is performed by a 32GS/s digitizer with a 10GHz analog bandwidth.

\subsection{Circuits}
\begin{figure}[H]
    \centering
    \includegraphics[width=0.95\columnwidth]{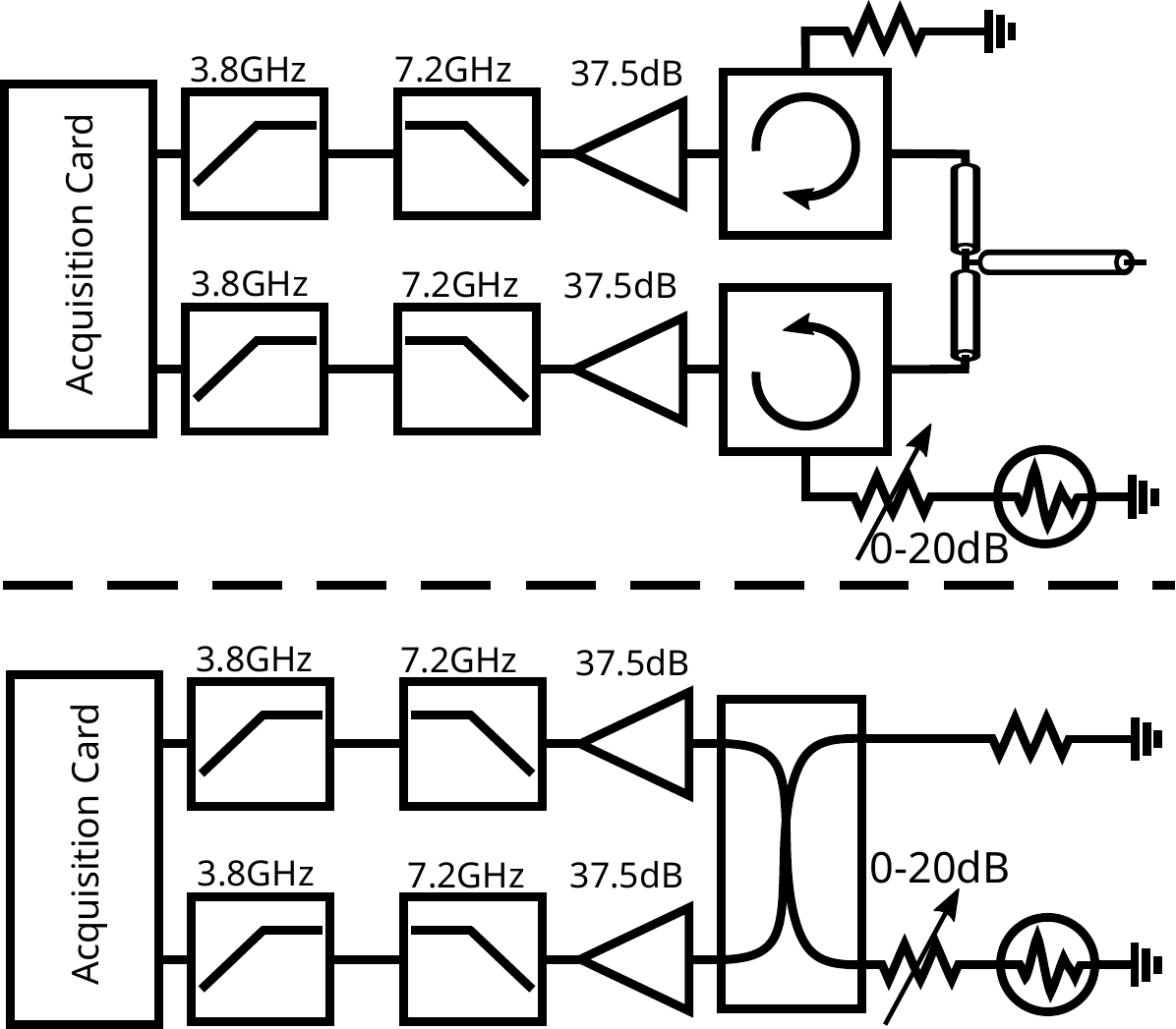}
    \caption{The setup consists of a room temperature resistor acting as a 
    cold noise source, 300K, and a calibrated noise source in series 
    with a variable attenuator which acts as a controllable hot noise 
    source with temperatures between 300K and 900K.
    Both are connected to the inputs of a 90$^\circ$ or 180$^\circ$ 
    hybrid coupler.
    The signals exiting the coupler are amplified with a gain of 37.5 dB 
    between with an amplifier noise temperature of 70K. 
    The signals are then filtered between 3.8GHz and 
    7.2GHz and are amplified again by the built-in amplifiers of 
    the digitizer before being digitized at a rate of 32 GS/s.}
    \label{fig:setup}
\end{figure}

We have studied three circuits. The first one is depicted on the top of Fig.~\ref{fig:setup} and is identical one 
of the circuits studied in \cite{dumont_violation_2023} (see Fig. 2a). 
It is made of several pieces of transmission lines. The two sources inject their noise into the circuit 
through circulators. 
The two noises experience many reflections and interferences within the circuit and leave via the two circulators. 
Thus the detected voltages are linear combinations of the two noises emitted at different times in the past, 
which in Fourier space reads:
\begin{align}
		V_1(f) &= S_{11}(f)\eta_1(f)+S_{12}(f)\eta_2(f),\label{eq:v1}\\
		V_2(f) &= S_{21}(f)\eta_1(f)+S_{22}(f)\eta_2(f)\label{eq:v2}.
\end{align}
\noindent where $\eta_{1,2}$ are the amplitudes of the signals generated by the noise sources at frequency $f$. 
The (scattering ~\cite{pozar_microwave_2011}) matrix $S_{ij}$ has a complex frequency dependence which reflects 
the various delays and interferences in the circuit, see Fig. 3(a1) in \cite{dumont_violation_2023}. 
The noise emitted by the detection chain is dumped into the noise sources thanks to the circulators, so it 
does not contribute to the measurement.

The two other circuits we have studied are a 180$^\circ$ and a 90$^\circ$ hybrid couplers, see Fig.~\ref{fig:setup} 
bottom. 
In contrast with the previous circuits, the response of the couplers is almost frequency independent within our 
detection bandwidth. Their scattering matrix are:
\begin{align}
    S_{\rm 180} = \frac{1}{\sqrt{2}}\begin{pmatrix}1&1\\-1&1\end{pmatrix},\quad
    S_{\rm 90} = \frac{1}{\sqrt{2}}\begin{pmatrix}1&i\\i&1\end{pmatrix},
\end{align}
where $i=\sqrt{-1}$. 

\section{Data acquisition and analysis}
The voltages $V_{1,2}$ are digitized 
simultaneously with 8 bits of resolution in batches of $2^{28}$ samples (512 MB per batch).
From the two time traces, we compute a 4D histogram of $[V_1(t),V_2(t),V_1(t+\tau),V_2(t+\tau)]$. 
This histogram, once normalized, provides a measurement of $P_\tau(\vec V,\vec V')$. From this we deduce
\begin{equation}
    P(\vec V)=\sum_{\vec V'}P_\tau(\vec V, \vec V')    
\end{equation}
and $j(\vec{V},\vec{V'})$ and $\vec{J}(\vec{V})$ 
according to Eqs.\eqref{eq:j} and \eqref{eq:vec_J}.

In order to perform the calculation of $P_\tau$ in a reasonable time, the data are split 
into 64 threads on a 36 core, 72 threads computer.
Each thread must have a local version of the histogram in RAM. 
With 8 bit resolution and 32 bit counters, this requires 16GB of RAM per thread, which is beyond the capabilities 
of our server. 
Thus we reduced the resolution of the measurement to 6 bit by bit shifting of the raw data and computed histograms 
with this resolution. 
A full measurement consists of 16 batches of 300 acquisitions of $2^{28}$ samples per channel.
Each batch is saved individually and consists of around 80 GS per channel and takes an hour to measure and compute, 
making the full measurement 1.2 TS per channel over 16 hours.

\section{Results}
\subsection{Transmission line}

\begin{figure}
    \centering
    \includegraphics[width=\columnwidth]{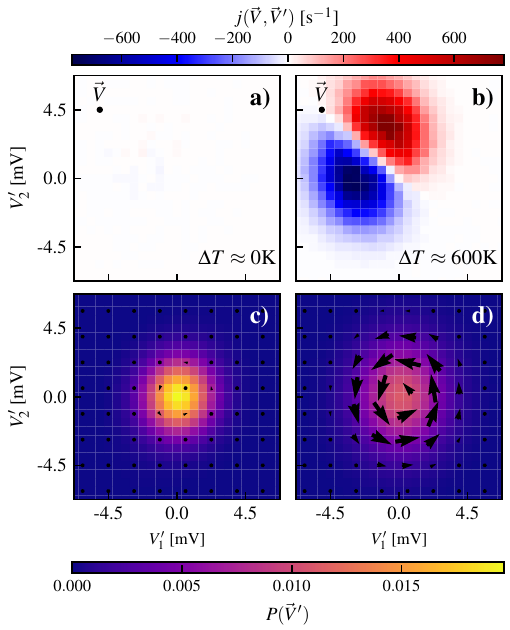}
    \caption{(Top) Probability current $j(\vec{V},\vec{V}')$ reconstructed for the 
    transmission line and circulators where point $i$ is marked by a black dot 
    and other pixels represent all possible points $j$ at $\Delta\approx 0$K 
    and $\Delta\approx 600$K respectively.
    (Bottom) Colormap representing the probability of points 
    $(V_1',V_2')$ measured for the transmission line and circulators at $\Delta\approx 0$K 
    and $\Delta\approx 600$K respectively. 
    The vector field represents the total probability current $\vec{J}(\vec{V}')$ at each 
    point.}
    \label{fig:detailed_balance_delay}
\end{figure}

We first consider the circuit made of transmission lines shown in Fig.\ref{fig:setup} top. 
The results of the measurement of $j(\vec{V},\vec{V}')$ for an arbitrary $\vec V$ are shown in 
Fig.\ref{fig:detailed_balance_delay})a for $\Delta T\approx 0$K and in Fig.\ref{fig:detailed_balance_delay})b 
for $\Delta T\approx 600$K. 
One observes that $j(\vec{V},\vec{V}')=0$ for all $\vec{V}'$ when $\Delta T=0$, i.e. detailed balance is 
obeyed at equilibrium. 
In contrast, Fig.\ref{fig:detailed_balance_delay})b shows a clear violation of the detailed balance when the two 
noise sources have different temperatures, i.e. when the circuit is driven out of equilibrium. 
There are two well defined regions of $\vec V'$ where $j(\vec{V},\vec{V}')\neq 0$. 
These regions become closer and shallower when $\Delta T$ is reduced (data not shown) and disappear for $\Delta T=0$. 
Note that the conservation of probability imposes
\begin{equation}
    \sum_{\vec V} j(\vec V,\vec V')=\sum_{\vec V'}j(\vec V,\vec V')=0
\end{equation}
i.e. the two regions of nonzero $j$ must have opposite sign, as observed. 
The existence of these regions has the following meaning: when the circuit is at position $\vec V$ in phase 
space at time $t$, it has a higher probability to be in the red region at time $t+\tau$ than to be in blue region. 
In the present circuit time reversal symmetry is obeyed, so $P_\tau(\vec V,\vec V')=P_\tau(\vec V',\vec V)$, 
which is experimentally verified with a very high precision. 
Thus the blue region also has the meaning of where the circuit was at time $t-\tau$. 
Having distinct points for where the system was at time $t-\tau$ and where it will be at time $t+\tau$ 
knowing where it is at time $t$ is the local expression of fluctuation 
loops~\cite{gonzalez_experimental_2019,ghanta_fluctuation_2017}.

The bottom part of Fig. \ref{fig:detailed_balance_delay} shows the probability distribution $P(\vec V)$ as 
background color and the vector field $\vec{J}(\vec{V})$ as arrows for $\Delta T=0$ (c) and $\Delta T=600$K (d). 
At equilibrium the variance of voltage fluctuations is smaller, hence the probability distribution is narrower 
and peaks at a higher value centered on $\vec V=\vec 0$. 
The current $\vec J$ vanishes at equilibrium and exhibits a vortex centered on $\vec 0$ that rotates 
counter-clockwise for $\Delta T\neq0$. 
$\vec J$ is zero in the center of the vortex, its magnitude shows a maximum for $|\vec V|$ of the order of the 
standard deviation of $\vec V$ and decays at large $|\vec V|$.
The counter-clockwise global rotation for the same circuit has been reported in \cite{dumont_violation_2023} as 
a finite, positive angular momentum, see Fig.4. The present measurement gives an insight of how the global 
rotation distributed in phase space. The existence of such a vortex is usual with out-of-equilibrium linear systems 
and has been reported in various implementations \cite{gonzalez_experimental_2019,chiang_electrical_2017}.

\subsection{Hybrid couplers}

\begin{figure}
    \centering
    \includegraphics[width=\columnwidth]{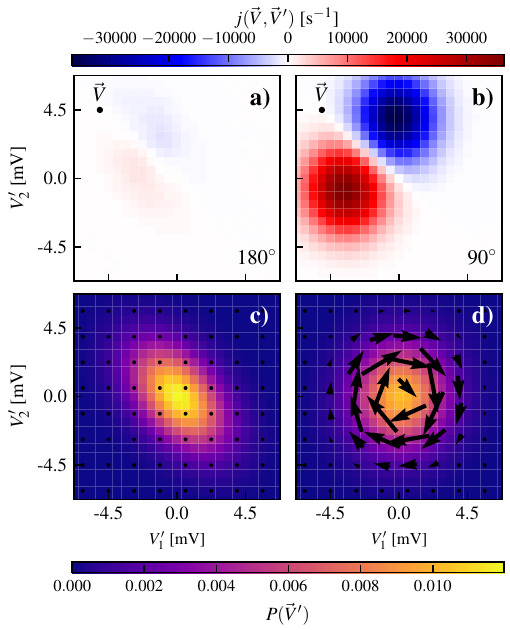}
    \caption{(Top) Probability current $j(\vec{V},\vec{V}')$ reconstructed for the 
    180$^\circ$ hybrid and 90$^\circ$ hybrid where point $i$ is marked by a black 
    dot and other pixels represent all possible points $j$.
    (Bottom) Colormap representing the probability of points 
    $(V_1',V_2')$ measured for the 180$^\circ$ hybrid and the 90$^\circ$ hybrid respectively. 
    The vector field represents the total probability current $\vec{J}(\vec{V}')$ at each 
    point.}
    \label{fig:detailed_balance_coupler}
\end{figure}

We now report the results for the hybrid couplers of Fig.\ref{fig:setup}(b). For both couplers we observe 
$j=0$ and $\vec J=\vec 0$ for $\Delta T=0$ (data not shown). 
The results obtained for $\Delta T=600$K are shown in  Fig.\ref{fig:detailed_balance_coupler}. 
The left column corresponds to the 180$^\circ$ coupler, the right one to the 90$^\circ$ coupler. 

The results are strikingly different: the probability current of the 180$^\circ$ hybrid is almost vanishing  
while that of the 90$^\circ$ hybrid is $\sim50$ times larger than the one of the circuit made of transmission lines. 
A simple explanation for this phenomenon has been given in~\cite{dumont_violation_2023} by considering the 
angular momentum in phase space: the rotation comes from each noise source contributing to the measured voltages 
with a dephasing, in a way similar to Lissajou curves on an oscilloscope. 
In the absence of dephasing, as with the 180$^\circ$ hybrid, the Lissajou curve collapses into a straight 
line with no rotation. 
Maximal rotation is obtained when the dephasing is 90 degree at each frequency, which corresponds to 
the 90$^\circ$ coupler. 
For any other circuit certain frequencies may contribute to clockwise rotation, others to anti-clockwise, 
resulting to an overall intermediate rotation, as for our circuit made of transmission lines.

The bottom part of Fig.\ref{fig:detailed_balance_coupler} shows $P(\vec{V})$ and the vector field $\vec{J}(\vec{V})$ 
for both couplers. 
While $P(\vec{V})$ for the 90$^\circ$ hybrid seems to have a rotational symmetry, as that of the first circuit, 
it is clearly elliptic for the 180$^\circ$ hybrid. 
The reason is the following: the scattering matrix $S_{180}$ corresponds to a 45-degree rotation in the 2D plane, 
so the probability distribution of $(V_1,V_2)$ is that of the noise sources, $(\eta_1,\eta_2)$, i.e. a Gaussian 
with variances proportional to $T_1$ and $T_2$, rotated by 45 degree. 
The ellipticity is a consequence of $\Delta T$ being nonzero. In contrast, if the scattering shows frequency 
dependence or is not real, $P(\vec V)$ is not simply related to $P(\eta_1,\eta_2)$ since the voltage at a given 
time involves $\eta_{1,2}$ in the past. 
This results in an apparent, but maybe not necessarily exact, rotational invariance of $P(\vec V)$.

\begin{figure}[H]
    \centering
    \includegraphics[width=\columnwidth]{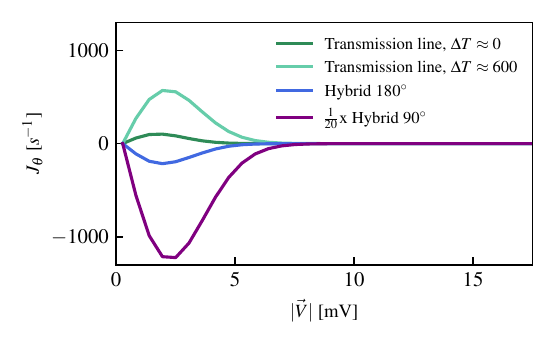}
    \caption{Angular component $J_\theta$ of the phase space current vector $\vec J$ as a 
    function of radius in voltage space, averaged for $\theta=0$ 
    and $\theta=\pi$.}
    \label{fig:J_V}
\end{figure}
The approach in frequency domain of \cite{dumont_violation_2023} allows to discriminate how different frequencies contribute to the overall rotation in phase space, measured by the total angular momentum. The present approach is dual, in the sense that it is blind to frequencies but probes the distribution of the current in phase space and not only the overall rotation. As a consequence, it allows to answer questions such as how much the rotation spreads in phase space. As an example we show in Fig.\ref{fig:J_V} how the orthoradial current, averaged for $\theta=0$ and $\theta=\pi$, varies as a function of the radius in voltage space. In order to be able to show all results on the same graph we scaled the data for the 90$^\circ$ hybrid by a factor  $1/20$. The case of the transmission line circuit and that of the 180$^\circ$ hybrid are not strictly zero most probably due to experimental imperfections. The shape of $J_\theta(|V|)$ shown in Fig.\ref{fig:J_V} has never been calculated as far as we know. The typical spread in voltage, here $\sim 2$mV for all curves should involve the two temperatures $T_{1,2}$ and not solely $\Delta T$. However the amplitude of $J_\theta$ should be proportional to $\Delta T$. 

\section{Conclusion and perspectives}
We have demonstrated the experimental reconstruction of the probability current in microwave circuits up to $\sim 10$GHz. 
In particular, we have observed local consequences of fluctuation loops. 
This study opens the route to that of similar circuits in the quantum regime where the energy associated with the 
frequency is smaller than the thermal energy, since 1 degree Kelvin corresponds to $\sim20$ GHz. 
In such a limit vacuum fluctuations play a fundamental role, which could be probed using our method.

\section*{Acknowledgments}
This work was supported by the Canada Research Chair program,the NSERC, the Canada First Research Excellence Fund, 
the FRQNT, and the Canada Foundation for Innovation.

\bibliographystyle{ieeetr}
\bibliography{ICNF_2023}

\end{document}